\documentclass[aps,prl,twocolumn,groupedaddress,showpacs]{revtex4}

\usepackage{graphicx}
\usepackage{dcolumn}
\usepackage{bm}
\usepackage{amsmath}

\newcommand{\s}{\ensuremath{\sigma}}
\newcommand{\te}{\ensuremath{\sigma}}

\begin{document}


\title{Atomic correlations in itinerant ferromagnets:
quasi-particle bands of nickel}

\author{J.~B\"unemann$^{1}$, F.~Gebhard$^{2}$, T.~Ohm$^{3}$, R.~Umst\"atter$^{3}$,
S.~Weiser$^{3}$, W.~Weber$^{3}$, \\
R.~Claessen$^{4}$, D.~Ehm$^{5}$, A.~Harasawa$^{6}$, A.~Kakizaki$^{6}$, 
A.~Kimura$^{7}$, G.~Nicolay$^{5}$, S.~Shin$^{6}$, and
V.N.~Strocov$^{4}$}
\affiliation{$^{1}$ Oxford University, Physical and 
Theoretical Chemistry Laboratory, 
Oxford OX1 3QZ, United Kingdom\\
$^{2}$ Fachbereich Physik, Philipps--Universit\"at Marburg, 
D--35032 Marburg, Germany\\
$^{3}$ Institut f\"ur Physik, Universit\"at Dortmund, D--44221 Dortmund, Germany\\
$^{4}$ Experimentalphysik II, Universit\"at Augsburg, D--86135 Augsburg, Germany\\
$^{5}$ Fachrichtung Experimentalphysik, Universit\"at des Saarlandes, D--66041
Saarbr\"ucken, Germany\\
$^{6}$ Institute for Solid-State Physics, 
The University of Tokyo, Kashiwa, Chiba 277-8581,
Japan\\
$^{7}$ \hbox{Dept.~of Physical Sciences, Graduate School of Science,
Hiroshima University, Higashi, Hiroshima 739-8526, Japan}
}


\begin{abstract}%
We measure the band structure of nickel along various high-symmetry
lines of the bulk Brillouin zone
with angle-resolved photoelectron spectroscopy.
The Gutzwiller theory for a nine-band Hubbard model
whose tight-binding parameters are obtained from non-magnetic
density-functional theory resolves most of the long-standing
discrepancies between experiment and theory on nickel.
Thereby we support the view of itinerant ferromagnetism as induced by 
atomic correlations.
\end{abstract}


\maketitle

\paragraph{Introduction.}

Studies of the electronic structure of ferromagnetic Ni, in particular
by angle-resolved photoelectron spectroscopy (ARPES),
have since long revealed large discrepancies between
experiments and results from spin-density-functional theory 
(SDFT)~\cite{Cal80,Mor78,LB23,Do94}.
The width of the
occupied part of the 3$d$ bands is approximately 
$W_{3d} = 3.3\, {\rm eV}$~\cite{EHK78,EP80},
whereas all SDFT calculations yield values of $W_{\rm SDFT} 
= 4.5\, {\rm eV}$~\cite{Mor78}.
In SDFT the exchange splitting of majority ($\uparrow$)
and minority ($\downarrow$) spin bands is
large and rather isotropic whereas the ARPES results report
small and highly anisotropic splittings. 
Typical is the problem of the $X_{2\downarrow}$ state which is positioned at
$0.04\, {\rm eV}$ below the Fermi energy~$E_{\rm F}$ by the ARPES data. 
In contrast, all SDFT calculations predict the
$X_{2\downarrow}$ state
to lie above~$E_{\rm F}$; so does a study using the GW approximation~\cite{Ary92}.
As a consequence,
a Fermi surface is predicted with {\sl two\/} hole ellipsoids
around the $X$~point whereas only one has been found in de Haas-van Alphen
experiments~\cite{Tsui67}, as later confirmed by ARPES~\cite{Do94,EHK78,EP80}.

Here, we present precise ARPES data on the electronic
structure of Ni, in which full control over the three-dimensional 
wave vector ${\mathbf k}$ was achieved using
very low-energy electron diffraction (VLEED) and ARPES~\cite{Stro98}. 
For comparison we use the quasi-particle bands
from Gutzwiller theory~\cite{Bue98,Thul02} which is based on 
a multi-band Hubbard model whose electron-transfer amplitudes are 
derived from non-magnetic DFT calculations.
We obtain good agreement with our ARPES data whereby we resolve
for the first time 
all of the qualitative discrepancies between experiment and SDFT
for the band structure of Ni.
We argue that the DFT calculations underestimate the ratio between the partial
charge densities in the 4$sp$ and the 3$d$ bands.
This appears to be a general
problem for DFT treatments of transition metals,
see Ref.~\onlinecite{note}.

\paragraph{Experiment.}

ARPES is a well-established tool for the measurement of
quasi-particle dispersions~$E({\mathbf k})$ in crystalline solids.
However, a principal problem concerns the determination of the
surface-perpendicular wave vector component~$k_{\perp}$ of the
photoelectron which is changed 
when the photoelectron exits into the vacuum.
Full control of the three-dimensional wave vector ${\mathbf k}$ {\em within} the
solid can be regained, if the dispersion of the photoelectron 
{\em final} states is known. The free-electron 
approximation 
which is usually employed for this 
purpose~\cite{EHK78,EP80,Kaki98} frequently fails
because the final-state bands are generally far more complicated~\cite{Stro98}.

A precise determination of both quasi-particle energies and
wave vectors is essential for a detailed comparison between
experiment and theory. To this end, we have 
controlled ${\mathbf k}$ using the following route.
First, angle-dependent VLEED is applied to determine the unoccupied
states above the vacuum level whose momentum ${\mathbf k}$ lies on surface-parallel
high-symmetry lines of the Brillouin zone (BZ), relying on the fact that
the photoelectron final states can be viewed as time-reversed LEED states.
Photoemission via these states, employing the Constant-Final-State (CFS) mode, 
is then used
to map out the dispersion of the occupied quasi-particle bands along
these lines. The usefulness and accuracy of this method has previously been
demonstrated for Cu~\cite{Stro98}. 

\begin{figure*}
\centerline{\includegraphics[height=10cm]{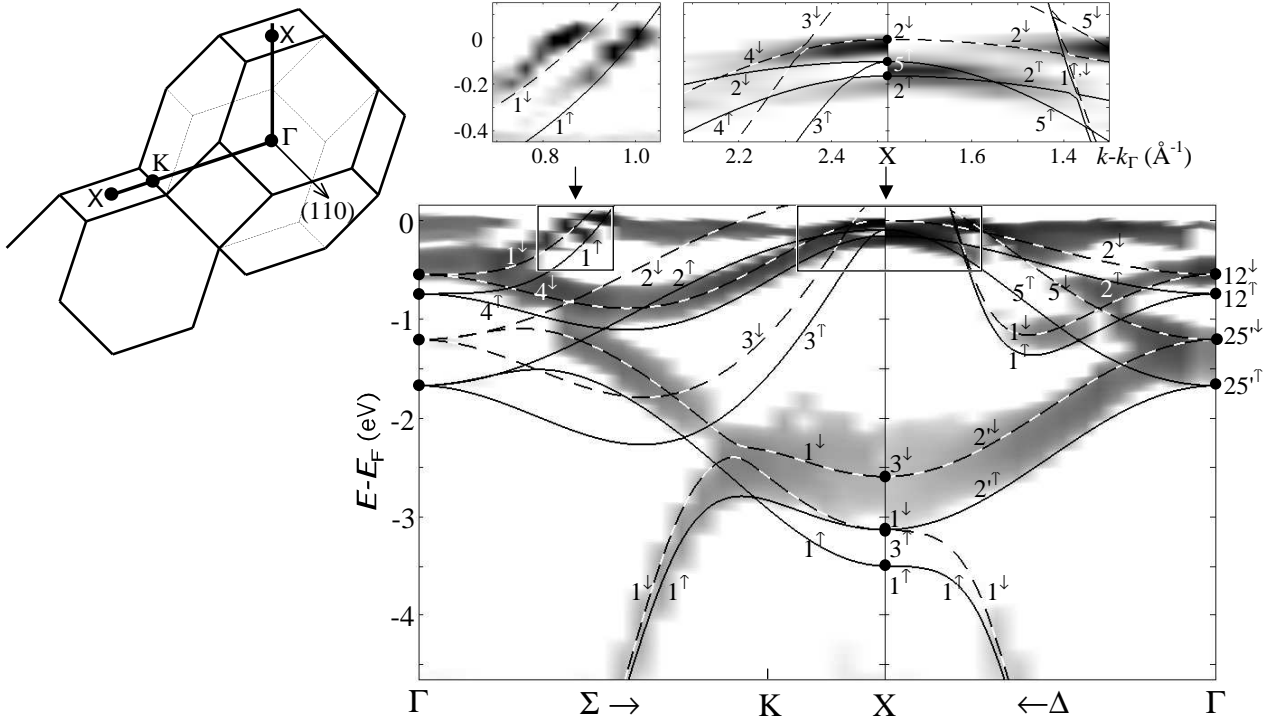}}
\caption{Grey-scale plot of the negative second derivative of the ARPES intensity
for nickel with respect to energy, $-d^2I/dE^2$, on a logarithmic scale (insets:
linear scale) for the $\Gamma KX$ and $\Gamma X$
directions of the BZ. The dispersionless structure at $E_F$ is due
to a residual Fermi edge resulting from indirect transitions. Some bands
($\Delta_2$, $X_1$, $\Sigma_2$, $\Sigma_3$,
$\Sigma_1$ towards $\Gamma$) are not seen
due to unfavorable matrix elements, depending on geometry
and chosen final state. Theoretical curves are G-DFT III, see
table~\protect\ref{table1}.
\label{fig1}}
\end{figure*}

In this work we report on measurements of the Ni(110) surface. The VLEED experiment
was performed with a conventional LEED unit operated in the retarding field
mode. ARPES in the CFS mode was carried out at beamline 18-A 
of the Photon Factory (Tsukuba, Japan);
see Ref.~\onlinecite{Stro98} for further details.
Both VLEED and ARPES measurements were performed along the 
$\overline{\Gamma}$$\overline{\rm X}$,
$\overline{\Gamma}$$\overline{\rm Y}$, 
and $\overline{\Gamma}$$\overline{\rm S}$ azimuths of the (110)
surface, giving access to the band dispersions along the $\Gamma$$K$$X$,
$\Gamma$$X$, and $X$$L$~lines of the bulk Brillouin zone. The ARPES data on the
quasi-particle dispersions are presented in Fig.~\ref{fig1} as a
gray-scale map of the negative second derivative of the 
photocurrent~\cite{Stro98}. Critical point energies are given in
Table~\ref{table1}.

\begin{table*}
\caption{Binding energies in~${\rm eV}$ with respect to the Fermi energy~$E_{\rm F}$
($>0$ for occupied states). $\langle \ldots \rangle$ indicates the spin
average, error bars in the experiments without spin resolution
are given as $\pm$. 
Theoretical data show the spin average and the exchange splittings in square
brackets. The second column denote the orbital character of the states,
$t\equiv t_{2g}$, $e\equiv e_g$,
capital letters: dominant character. The spin-polarized data 
$\langle \Lambda_{3;f}\rangle$
were taken at fractions~$f$ of the $\Gamma$$L$ distance,
with the emphasis on the analysis of the exchange splittings.
For all calculations, including the SDFT, a spin-only moment
of $\mu=0.55 \mu_{\rm B}$ was used.
G-DFT~I: $A=12\, {\rm eV}$, $n_d=8.8727$;
G-DFT~II: $A=9\, {\rm eV}$, $n_d=8.8727$;
G-DFT~III: $A=9\, {\rm eV}$, $n_d=8.7793$.
\label{table1}}
\begin{ruledtabular}
\begin{tabular}{lclcllll}
 Symmetry &     Character &    
Experiment & Reference & G-DFT I & G-DFT II & G-DFT III &  SDFT\\
\hline
\mbox{}&&&&&&&\\[-6pt]
$\langle \Gamma_1 \rangle$ & S &  8.90$\pm$0.30 &&     8.87 &  8.89 & 8.86 
& 8.96[$-$0.11]\\
$\langle \Gamma_{25'}\rangle$& T& 1.30$\pm$0.06 && 1.34[0.62]& 1.52[0.62]& 1.44[0.46]
& 1.99[0.43]\\
$\langle \Gamma_{12}\rangle$ & E & 0.48$\pm$0.08 && 0.62[0.28]& 0.71[0.29] & 0.65[0.195] 
& 0.86[0.41]\\
$\langle X_1 \rangle$        & sE& 3.30$\pm$0.20 &
\protect\cite{EP80}
& 2.98[0.57]& 3.39[0.50]& 3.31[0.36]
& 4.37[0.20]\\
$\langle X_3 \rangle$        & T & 2.63$\pm$0.10 &&   2.60[0.78]& 2.95[0.72]& 2.86[0.54]
&3.82[0.37]\\
$ X_{2\uparrow}$         & E &   0.21$\pm$0.03 &&  0.23    &  0.27    &   0.165 
& 0.35\\
$ X_{2\downarrow}$                 & E  &  0.04$\pm$0.03 &&   0.02    &  0.02   &    0.01
& $-$0.09\\
$ X_{5\uparrow}$                &  T &  0.15$\pm$0.03  &&   0.22   &    0.24  &     0.10
& 0.23\\
$\Delta_{e_g}(X_2)$                &  E &   0.17$\pm$0.05      
&&   0.214   &    0.25  &     0.155
&  0.44\\
$\Delta_{t_{2g}}(X_5)$            &   T &   0.33$\pm$0.04      &
\protect\cite{EHK78}&  0.45 &     0.49 &      0.38 
&  0.56\\
%
%
$\langle K_2 \rangle$   & spTe& 2.48$\pm$0.06 &&     2.33[0.70] &  2.64[0.66]&  2.59[0.50]
&   3.37[0.32]\\
%
%
$\langle K_4 \rangle$   & pE&   0.47$\pm$0.03 &&     0.51[0.26] &  0.58[0.28] & 0.51[0.185]
&  0.70[0.41]\\
%
%
$\langle L_1 \rangle$  &  sT&   3.66$\pm$0.10 &&     3.21[0.76]  & 3.59[0.68]    
&3.51[0.515] &  4.56[0.23]\\
$\langle L_3 \rangle$  &  tE&   1.43$\pm$0.07 &&     1.39[0.48]  & 1.58[0.47]    
&1.51[0.34] &   2.02[0.40]\\
$ L_{3\uparrow}$        &   Te&   0.18$\pm$0.03 &&     0.30[0.37]&  0.34[0.41] 
&0.215[0.30] &  0.38[0.50]\\
$\langle L_{2'}\rangle$&    P&    1.00$\pm$0.20 &
\protect\cite{EP80}
&   0.32     &   0.29[0.0]  & 0.97[0.0]
&    0.24[$-$0.12]\\
%
%
%
$\langle W_1 \rangle$&    sE &  0.65$\pm$0.10 &
\protect\cite{EP80}
&   0.66[0.28]  & 0.76[0.30]  
& 0.69[0.20] &  0.94[0.39]\\
$ W_{1'\uparrow}$&   T     & 0.15$\pm$0.10 &
\protect\cite{EP80}
&   0.23[0.46]  &0.25[0.52] & 0.11[0.38]  
&0.23[0.56]\\
$\langle \Lambda_{3;1/3}\rangle$
& ptE&  0.57[0.16$\pm$0.02]
&
\protect\cite{Kaem90}
&  0.64[0.30] &  0.73[0.32] & 0.67[0.22]
&   0.90[0.42]\\
$\langle \Lambda_{3;1/2}\rangle$
& ptE&  0.50[0.21$\pm$0.02]
&\protect\cite{Kaem90}
&  0.54[0.34] &  0.61[0.37] & 0.55[0.26]
&   0.76[0.44]\\
$\langle \Lambda_{3;2/3}\rangle$
& pTE&  0.35[0.25$\pm$0.02]
&\protect\cite{Kaem90}
&  0.35[0.37] &  0.39[0.41] & 0.33[0.29]
&   0.49[0.48]\\
\end{tabular}
\end{ruledtabular}
\end{table*}

\paragraph{Gutzwiller theory.}

The Gutzwiller theory employs a tight-binding Hamiltonian for 
a basis of 4$s$, 4$p$ and 3$d$ orbitals
\begin{equation}
\widehat{H}_1=\sum_{i\neq j=1;\te,\te'}^{L}
t_{i,\te;j,\te'}
\widehat{c}_{i;\te}^{\, +}
\widehat{c}_{j;\te'}^{\vphantom{\, +}}
+ \sum_{i;\te}
\widetilde{\epsilon}_{\te} \widehat{c}_{i;\te}^{\, +}
\widehat{c}_{i;\te}^{\vphantom{\, +}}
 \; .
\label{Hitinerant}
\end{equation}
The energy-transfer integrals
$t_{i,\te;j,\te'}$
between the spin-orbitals~$\te$ at site~$i$ and $\te'$ at site~$j$
range up to third nearest neighbors. 
They result from a fit to DFT energy bands of non-magnetic Ni, calculated
with the LAPW-WIEN code~\cite{WIENXX} and using the local-density approximation;
for details of the fit procedure, see Ref.~\onlinecite{WM82}.
A value of $15\, {\rm meV}$ for the root-mean-square deviation 
was obtained for all bands up to $2\, {\rm eV}$ above the Fermi energy.
We note that the DFT energy of the 4$p$-type $L_{2'}$ state 
is $0.120\, {\rm eV}$ below $E_{\rm F}$, and the bare 3$d$ band width in our
tight-binding model is $W_{\rm bare}\approx W_{\rm DFT}=
4.5\, {\rm eV}$.

The single-particle term is supplemented by the atomic Coulomb
interactions among the 3$d$ electrons,
\begin{eqnarray}
\widehat{H}&=&\widehat{H}_1 + \sum_i\widehat{H}_{i;{\rm at}} \; ,
\label{Htotal}
\\
\widehat{H}_{i;{\rm at}}
&=&
\sum_{\s_1,\s_2,\s_3,\s_4}
{\cal U}^{\s_1,\s_2;\s_3,\s_4}
\widehat{c}_{i;\s_1}^{\, +}\widehat{c}_{i;\s_2}^{\, +}
\widehat{c}_{i;\s_3}^{\vphantom{\, +}}
\widehat{c}_{i;\s_4}^{\vphantom{\, +}} 
\nonumber \\
&=& 
\sum_{\Gamma} E_{\Gamma}\widehat{m}_{i;\Gamma}
\;.
\label{Hat}
\end{eqnarray}
Here, $\widehat{m}_{i;\Gamma}$ is the projector onto the atomic eigenstate
$|\Gamma\rangle_i$ with energy~$E_{\Gamma}$. 
In the spherical approximation, the Coulomb 
parameters~${\cal U}^{\s_1,\s_2;\s_3,\s_4}$ can be obtained
from three Racah parameters. In this work we choose $B = 0.09\, {\rm eV}$ 
and $C = 0.40\, {\rm eV}$,
guided by spectroscopic data for isolated Ni$^{2+}$ and Ni$^{3+}$ ions;
see Table~5.1 of Sugano et al.~\cite{Su70}. This choice fulfills the
`canonical' ratio $C/B \approx 4.5$, found 
for transition metal ions both in atomic and impurity spectroscopy~\cite{Su70}.
The Racah parameter~$A={\cal O}(10\, {\rm eV})$ is adjusted; see below.

The sizable atomic Coulomb interactions suppress 
local charge fluctuations~\cite{vV53} which occur in the limit of
independent particles. Therefore, Gutzwiller~\cite{Gu63}
proposed to address the states
\begin{equation}
|\Psi_{\rm G}\rangle =\widehat{P}_{\rm G}|\Phi\rangle \quad , \quad
\widehat{P}_{\rm G} =
\prod_{i;\Gamma} \eta_{\Gamma }^{\widehat{m}_{i;\Gamma}}
\label{DefGutzwiller}
\end{equation}
as a variational approximation for the true ground state of $\widehat{H}$.
The real numbers~$\eta_{\Gamma }$ parameterize
the Gutzwiller correlator~$\widehat{P}_{\rm G}$
which reduces energetically 
unfavorable atomic configurations in
the normalized one-particle product state $|\Phi\rangle$.
For an open $d$-shell system, there are $2^{10}$, i.e.,
of the order of $10^3$ variational parameters.

The variational ground state is determined by a minimization 
of the expectation value $\langle \widehat{H} \rangle_{\Psi_{\rm G}}$ 
with respect to the `internal' parameters $\eta_{\Gamma }$ 
and the wave function $|\Phi\rangle$. 
In infinite dimensions~\cite{Bue98,Thul02} this leads to the 
condition that  $|\Phi\rangle$ is the ground state of the effective 
Hamiltonian 
\begin{eqnarray}
\widehat{H}^{\, \rm eff}&=& 
\sum_{{\mathbf k}, \sigma,\sigma^{\prime}}\widetilde{S}_{\te,\te'}({\mathbf k})
\widehat{c}_{{\mathbf k},\te }^{\, +}\widehat{c}_{{\mathbf k},\te' }^{}    
\label{qp-dispersion} \; ,\\
\widetilde{S}_{\te,\te'}({\mathbf k}) &=&  
\sqrt{q_{\te}} \sqrt{q_{\te'}} \epsilon_{\te,\te'}({\mathbf k}) +    
\delta_{\te,\te'} (\widetilde{\epsilon}_{\te}+\lambda_{\te}) \; ,  
\label{Heffb} 
\end{eqnarray} 
where $\epsilon_{\te,\te'}({\mathbf k})$
is the Fourier transform of the 
bare electron-transfer integrals 
$t_{i,\te;j,\te'}$.
The spin-orbital energy shifts $\lambda_{\te}$ act as `external' 
variational parameters since they determine $|\Phi\rangle$ 
via the eigenvalue equation $(\widehat{H}^{\, \rm eff}-E)|\Phi\rangle=0$. 

The quasi-particle excitations from energy bands given by
the eigenvalues $E({\mathbf k},\gamma)$ of 
$\widehat{H}^{\, \rm eff}$~\cite{Thul02}. 
Note, that the effective 
single-particle 
Hamiltonian differs from 
$\widehat{H}_{1}$ by the renormalization factors  
$0\leq q_{\te}\leq 1$ and the energy shifts $\lambda_{\te}$. 
Therefore, the bare bands of $\widehat{H}_{1}$ become narrowed and mixed 
in the quasi-particle excitations of the correlated system.
The above results are strictly valid only for $D=\infty$. 
However, $1/D$~corrections were found to be small for $D=3$ 
single-band Fermi-liquid systems~\cite{Thul02}.  

The Gutzwiller method overestimates the optimum magnetic moment 
by about 10\% to 15\%; for a remedy of this
problem, see Ref.~\onlinecite{Ohm02}.
Therefore, we work with a fixed magnetic moment~$\mu=0.55\mu_{\rm B}$, 
the experimental spin-only moment. 
Spin-orbit coupling has not yet been
incorporated into our Gutzwiller codes as this requires a major
reprogramming effort. 
We keep the partial densities
$n_s$, $n_p$, and $n_d$ fixed to the values
obtained from the DFT one-particle Hamiltonian.
We have studied other variants of applying 
the Gutzwiller-DFT 
which allow charge flow
between 4$sp$ and 3$d$ channels,
yet the effects on the quasi-particle 
bands are found to be small~\cite{Ohm02}. 

\paragraph{Results for nickel.}

In cubic iron-group metals and under the limitations of keeping $n_d$, $n_s$,
and $n_p$ fixed, symmetry allows for
three independent energy shifts $\lambda_{\sigma}$
for the 3$d$ electrons. They govern 
the exchange splittings 
$\Delta(t_{2g})$ and
$\Delta(e_g)$,
and the crystal-field splitting
$\Delta(e_g/t_{2g})$.
For Ni, the exchange splittings 
of the 4$s$ and 4$p$ orbitals turn out to be of very minor importance.
Therefore, we minimize the variational ground-state energy 
for a fixed magnetic moment~$\mu$
with respect to three external 
and about 400 internal variational parameters.

We find the best agreement with the occupied 3$d$ band width $W_{3d}=3.3\, {\rm eV}$
at $X_1$ for a value $A = 9\, {\rm eV}$, see table~\ref{table1}, 
column~G-DFT~II. 
Then, the condensation energy, i.e., the
energy gain for the ferromagnetic phase as compared to the paramagnetic one,
is $E_{\rm cond} \approx 40\, {\rm meV}=0.8 k_{\rm B}T_{\rm C}$
($T_{\rm C}=630\, {\rm K}$ is the Curie temperature).
This value of~$A$ is much larger than in other multi-band studies~\cite{Pru97,Kot01}.
In~\cite{Pru97}, only 3$d$ bands were incorporated. If we eliminate 
in our calculations the rather
large 4$sp$-3$d$ hybridizations we also find that values of 
$A = 2\, {\rm eV}\ldots 3\, {\rm eV}$ 
give ferromagnetic solutions with 
$E_{\rm cond}=40\, {\rm meV}$ and $W_{3d}\approx 3\, {\rm eV}$.
The results of~\cite{Kot01}, where the 4$sp$ bands have also been included,
are at 
variance with our findings.

For all three G-DFT calculations of table~\ref{table1}, 
the energy of the state $X_{2\downarrow}$ 
is pinned slightly below the Fermi energy,
in agreement with experiment.
This is an essential improvement over SDFT and even SDFT-GW
calculations~\cite{Ary92}
which predict a second hole pocket at the $X$~point.
Even a drastic change in the
Racah parameter $A$ from $A_{\rm II}= A_{\rm III} =9\, {\rm eV}$ 
to $A_{\rm I}= 12\, {\rm eV}$
does not change much the basic features
of the Gutzwiller quasi-particle band structure.
The pinning of $X_{2\downarrow}$ is caused by the 
small variational parameter $\Delta(e_g)$,
which in turn leads to a large fraction of $t_{2g}$ holes in the minority spin
bands, produced by a relatively large $\Delta(t_{2g})$.  
Because of the large nearest-neighbor hopping between 
$t_{2g}$ orbitals, it is energetically favorable to generate
as many $t_{2g}$ holes in the
minority spin bands as possible.

A more detailed inspection of the table~\ref{table1}
indicates several significant discrepancies between experiment 
and the results from G-DFT~II. Most notably,
the energy of the state $L_{2'}$,
$E_{\rm GII}(L_{2'}) =-0.3\, {\rm eV}$, is much higher
than in experiment, $E(L_{2'}) =-1.0\, {\rm eV}$. 
This discrepancy is caused by the underlying DFT calculation
(see also table~\ref{table1}, column~SDFT),
as the 4$p$-type $L_{2'}$ state
is basically not affected by correlations in the 3$d$ shell.
When we lower the bare 4$p$~level $\epsilon_{p}$ in~(\ref{Hitinerant}) 
by $0.75\, {\rm eV}$,
we considerably improve the agreement with experiment
especially for all states near the Fermi energy
and thus also for the values of the exchange splittings.
The photoelectron data reveal
$\Delta_{e_g}(X_2)=170\pm 50\, {\rm meV}$ 
for the pure $d(e_g)$ state $X_2$, 
and $\Delta_{e_g}(X_5)=330\pm 40\, {\rm meV}$~\cite{EHK78} for the pure 
$d(t_{2g})$ state $X_5$, which agrees well with
the data from the G-DFT~III.
Lowering the 4$p$
level leads to a charge flow of about $0.1$~electrons from 3$d$ to 4$p$
whereby the number of
holes in the 3$d$ bands is 
significantly enhanced. Thus, there are more 3$d$ holes available to generate
the magnetic moment, and the exchange splittings
are reduced accordingly.
Note that the G-DFT exchange splittings increase as a function of the
binding energy by about 50\%, as can be seen, e.g.,
from the values of the pure $t_{2g}$ states $X_5$, $\Gamma_{25'}$, and
$X_3$. In contrast, they decrease in the SDFT because of the bigger
orbital basis used there.

\paragraph{Conclusions.}

Hartree--Fock theory was the starting point for the
Stoner--Slater band theories of itinerant ferromagnetism~\cite{Sla36,Sto38}. 
For decades these theories
have been the only ones to provide detailed comparison with
experiments.
In this paper we have shown for the prototypical ferromagnet Ni that,
starting from a DFT-based one-particle Hamiltonian, the
Gutzwiller theory can resolve the main discrepancies for Ni.
We thereby confirm the qualitative findings of previous model
studies~\cite{Bue98,Pru97} on the basic mechanism of itinerant
ferromagnetism. In Ni the 3$d^8$, 3$d^9$, and 3$d^{10}$ multiplets are 
predominantly occupied to accommodate the 3$d$ electrons' itinerancy.
However, the occupation of the low-lying 3$d^8$ spin triplets is
significantly enhanced over the occupation of the
3$d^8$ singlets. As in 
magnetic insulators, Hund's first rule 
is found to be an important agent in itinerant ferromagnets.
Therefore, we corroborate early ideas 
by van Vleck~\cite{vV53}, Gutzwiller~\cite{Gu63}
and others on the origin of itinerant ferromagnetism. In contrast to band aspects,
they emphasized the importance 
of local moments in narrow-band metals
as a consequence of strong atomic correlations.

Further predictions of our theory, e.g., 
the increase of the exchange splittings as a function of energy,
should be tested by more detailed spin-resolved experiments.

\begin{acknowledgments}
W.W.\ thanks S.G.~Louie and M.L.~Cohen for their hospitality
during his sabbatical stay.
We thank H.\ Starnberg and P.\ Blaha for their help
with the VLEED data acquisition and analysis.
This work was supported by the Deutsche Forschungsgemeinschaft 
(446JAP/113/104/0, BU1309/2-1, CL124/5-1, WE1412/8-1).
\end{acknowledgments}

\end{document}